\documentclass[]{agujournal}

\journalname{JGR-Space Physics}

\begin{document}

%
%

\title{Numerical Algorithm for Detecting Ion Diffusion Regions in the Geomagnetic Tail with Applications to MMS Tail Season May 1-- September 30, 2017}

%
%

 \authors{A. Rogers\affil{1}\thanks{UNH, NH},
~C.~J~Farrugia\affil{1}, and 
~R.~B.~Torbert\affil{1}
}
\affiliation{1}{Institute for the Study of Earth, Oceans, and Space, Dept of Physics, University of New Hampshire, Durham, NH}


\correspondingauthor{Anthony Rogers}{ajr1022@wildcats.unh.edu}


\begin{keypoints}
\item  Present a stepwise scheme for identifying IDRs in the near-Earth geomagnetic tail implemented using a numerical algorithm
\item  The algorithm finds 12 IDRs in the 2017 MMS geomagnetic tail season (May 1- September 30).
\item  Statistical properties of these IDRs are compared with those observed by Cluster in 2001-2005.

\end{keypoints}

%
%
\begin{abstract}

	We present a numerical algorithm aimed at identifying ion diffusion regions (IDRs) in the geomagnetic tail, and test its applicability. We use 5 criteria applied in three stages. (i) Correlated reversals (within 90 s) of V$_{x}$  and B$_{z}$ (at least 2 nT about zero; GSM coordinates); (ii) Detection of Hall electric and magnetic field signatures; and (iii) strong ($\geq 10 mV/m$) electric fields. While no criterion alone is necessary and sufficient, the approach does provide a robust, if conservative, list of IDRs. We use data from the Magnetospheric Multiscale Mission (MMS) spacecraft during a 5-month period (May 1 to September 30, 2017) of near-tail orbits during the declining phase of the solar cycle.  We find 148 events satisfying step 1, 37 satisfying steps 1 and 2, and 17 satisfying all three, of which 12 are confirmed as IDRs. All IDRs were within the X-range [-24, -15] RE mainly on the dusk sector and the majority occurred during traversals of a tailward-moving X-line.  11 of 12 IDRs were on the dusk-side despite approximately equal residence time in both the pre- and post-midnight sectors (56.5\% dusk vs 43.5\% dawn). MMS could identify signatures of 4 quadrants of the Hall B-structure in 3 events and 3 quadrants in 7 of the remaining 12 confirmed IDRs identified.  The events we report commonly display V$_{x}$ reversals greater than 400 km/s in magnitude, normal magnetic field reversals often >10 nT in magnitude, maximum DC |$\vec{E}$| which are often well in excess of the threshold for stage 3. Our results are then compared with the set of IDRs identified by visual examination from Cluster in the years 2000-2005.

\end{abstract}

%
%

%



%
%
%


\section{Introduction}
\label{sec:intro}

Reconnection is aprocess by which electromagnetic energy is transferred to the plasma, heating and accelerating the particles.  In a magnetospheric context, this process occurs on the dayside, where closed magnetic flux is opened, and on the nightside, where open flux is closed and transported back to the dayside. In an open magnetopshere, it thus drives the magnetospheric convection, as initially postulated by Dungey (1961).

Near the reconnection site, the ions and electrons become decoupled from the magnetic field. The breaking of the frozen-in condition takes place in two stages. First, the ions are decoupled from the magnetic field in what is called the ion diffusion region (IDR). Later, the electrons decouple in a much smaller region, appropriately called the electron diffusion region (EDR), where the magnetic fields also reconnect and change topology i.e. from closed to open and vice versa (Vasyliunas, 1975; Sonnerup, 1979). Thus diffusion regions are central to our understanding of energy conversion through reconnection. Hence the need to be able to properly identify them. Our focus here will be on IDRs.

Thus far, IDRs have been identified by visual inspection of the data. This has been done for large surveys of Geotail data (Nagai et al. 1998, 2005) and Cluster (Eastwood et al., 2010).  Several important properties came to light, which enhanced our understanding of these important regions.  Usually, visual identification depends on the observation of correlated reversals of the Sun-Earth component ($\hat{x}$) of the flow velocity and the north-south ($\hat{z}$) component of the magnetic field in proper coordinates. This reversal is taken as a sign of the X-line moving past the spacecraft. The IDR identification is then further strengthened when Hall magnetic and electric fields are identified as predicted by theory (Sonnerup, 1979) and notably observed by Øieroset et al. (2001). These fields arise from the differential motion of the electrons and ions.  In the tail, which is the region we direct our attention to here, the Hall magnetic fields typically form a quadrupolar structure pointing in the out-of-plane direction, while the Hall electric fields are bipolar and point towards the current sheet (Mozer et al. 2002, Wygant et al. 2005, Borg et al. 2005).  This is illustrated in Figure 1.  A strong electric field often accompanies observations of IDRs (Eastwood et al., 2010). 

While very useful, this visual identification can be tedious and prone to error.  For example, there are cases where establishing an association of oppositely-directed ion flows and Hall fields is problematic (Nagai et al. 2013). Clearly, a significant advance would be achieved if we were to speed up this process, eliminate some uncertainties, and improve consistency.

This is what we intend to do here. We develop a numerical algorithm to search for IDRs. The process has three stages. Only those events which satisfy all three are considered to be \textit{bona fide} IDRs. We then apply this to observations by the Magnetospheric Multi-Scale (MMS) spacecraft during the 2017 tail season from May 1 to September 30, spanning all the near tail and going from dawn to dusk from MLT 4:47 to 18:00 hrs MLT at low MLATs in 53 orbits. The statistical properties of the IDRs that the code finds, such as typical Hall magnetic and electric fields will be compared with those of Cluster as reported by Eastwood et al (2010).


\begin{figure}[h!]
	        \centering
	        \includegraphics[width=0.9\textwidth]{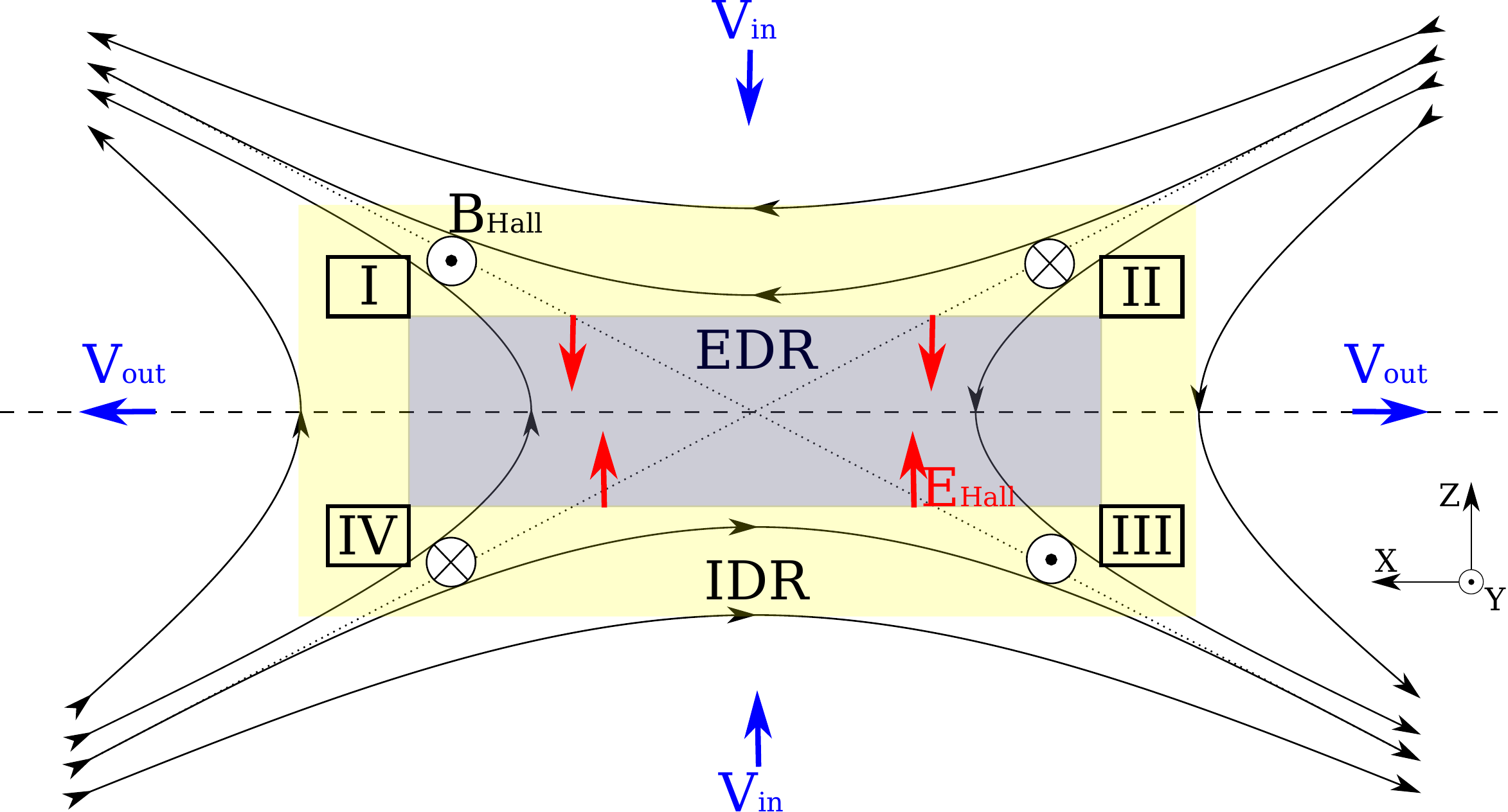}
	        \caption{An ideal picture of symmetric Hall reconnection showing the quadrupolar out-of-plane ($\hat{y}$ in this view) idealized magnetic fields and inflow and outflow regions in and surrounding the Ion Diffusion Region.}
	        \label{fig:xline}
\end{figure}

The geomagnetic tail is a preferred region for testing this algorithm. The reconnecting fields are approximately anti-parallel and, further, the field and plasma properties in the lobes above and below the current sheet are typically about the same so that the reconnection is under symmetric conditions. This mitigates complications arising from  the presence of a guide field and asymmetries in the inflow regions which do not let us know beforehand what Hall structures, such as the quadrupolar B-field structure, to expect (see e.g. Eastwood et al., 2013; Pritchett and Mozer, 2008, Muzamil et al., 2014, and references therein) since the asymmetries can alter them severely.  (See, however, Zhou, et al. (2016) who describe quadrupolar B fields under asymmetric conditions.)  Further, use of geocentric solar magnetospheric (GSM) coordinates is often good enough so that we do not need to transform to boundary-normal coordinates.  Only rarely does reconnection on the dayside approach symmetric conditions (see Mozer et al., 2002).

The layout of the paper is as follows.  In section 2 we describe the data we shall use, in section 3 we detail the Methodology and Procedure.  Section 4 gives two case studies, section 5 gives an overview of the output of the code as applied to the 2017 MMS tail season, section 6 shows statistical results over the ensemble of IDRs and Section 7 gives a discussion.  We compare various aspects of our findings with the works of Eastwood et al. (2010) and Nagai et al.(2005)   We shall also compare our findings with the predictions on the occurrence frequency of IDRs postulated by Genstreti et al. (2014). In section 8 we draw our conclusions.

\section{Instruments}
\label{sec:instr}

The method we describe utilizes magnetic and electric field and ion velocity moment data collected by the MMS spacecraft and published on the MMS science gateway website (https://lasp.colorado.edu/mms/sdc/public/).  These data were acquired by the FIELDS instrument suite in the case of magnetic and electric fields and by the Fast Plasma Investigation (FPI) for ion moments.

The MMS spacecraft measure electric and magnetic fields using the FIELDS instrument suite. FIELDS consists of three electric field and three magnetic field instruments (Torbert et al., 2016). The two pairs of spin-plane double probes (SDP) and the axial double probes (ADP) allow MMS to make direct measurements of the full 3D electric field, ranging from DC up to 100 kHz (Lindqvist et al., 2016, Ergun et al., 2016).  These data are combined into the EDP data product for 3D vector $\vec{E}$ measurements.  Version 3.0.0 of the level 2 EDP data products was used throughout this study.  The analog and digital fluxgate magnetometers (AFG/DFG) measure magnetic fields in the frequency range from DC up to 64 Hz (Russell et al., 2016). The higher frequency range, from 1 Hz up to 6 kHz, is covered by a search-coil magnetometer (SCM; Le Contel et al., 2016).  Level 2 fluxgate magnetometer (FGM) data of version 5.86 and higher (highest available as of submission) were used throughout this study.

The FPI provides MMS with high cadence electron and ion distributions in the energy/charge range of 10 eV/q up to 30 keV/q. Each MMS satellite is equipped with eight FPI spectrometers which, combined with electrostatic control of the field-of-view, allows FPI to sample in burst mode the full electron and ion distributions with a time resolution of 30 ms and 150 ms, respectively (Pollock et al. 2016). It is important to note that core ion distributions can extend beyond the range of FPI, meaning that actual ion bulk velocities may be higher than what is calculated using FPI data.  Identification of IDRs in our algorithm uses fast survey data.  Level 2 FPI ion moments of version 3.3.0 were used throughout this study.

\section{Methodology and Procedure}  

\begin{figure}[h!]
	\centering
	\includegraphics[width=\textwidth]{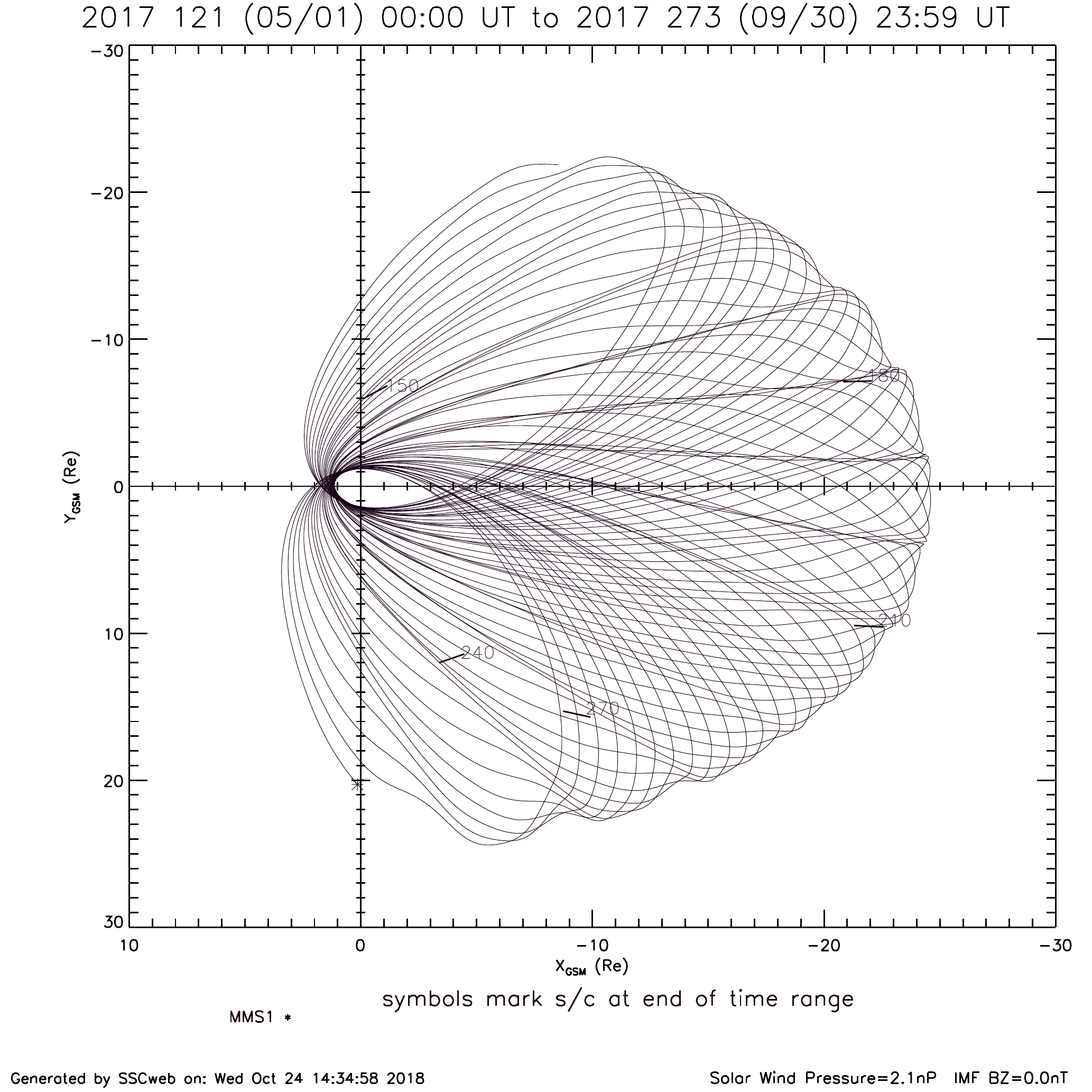}
	\caption{MMS coverage of the near-tail for the 5-month period of our study. The GSM Z was within -7 to 7 RE.  This shows a fairly complete coverage of the near-tail region during the period of our study.}
	\label{fig:orbit}
\end{figure}

Data collected by the MMS fleet of spacecraft were analyzed for the 5-month period from 01 May 2017 through 30 September 2017.  During this time the apogee of the spacecraft orbits reached a typical distance of $\sim 25R_{E}$ and swept from $\sim 5MLT$ to $\sim 19MLT$.  Orbital tracks are given in Figure \ref{fig:orbit}.  Central to us is the time the MMS spacecraft spends in the plasma sheet.  We adopt a criterion based on plasma density ($n \geq 0.05 \frac{1}{cc}$) which is taken as indicative of the plasma sheet (Baumjohann 1993, Raj et al. 2002).  Only time segments which meet this condition are considered for further analysis.  Dwell times for the spacecraft in the plasma sheet are represented as a part of Figure \ref{fig:dwell-xy}.

MMS data was analyzed in three-minute segments with a 1-minute overlap between adjacent segments.  Each segment determined to be in the plasma sheet was analyzed using a maximum of five search criteria: (i) Ion flow reversal in the GSM$_{x}$, Earth-Sun direction; (ii) magnetic field reversal in the GSM$_{z}$ direction; (iii) sign correlation with the flow reversal; (iv) presence of Hall electric, and Hall magnetic fields; and (v) magnitude of the measured electric field.  These criteria were applied to each three-minute segment in sequential stages.

\begin{figure}[h!]
	\centering
	\includegraphics[width=\textwidth]{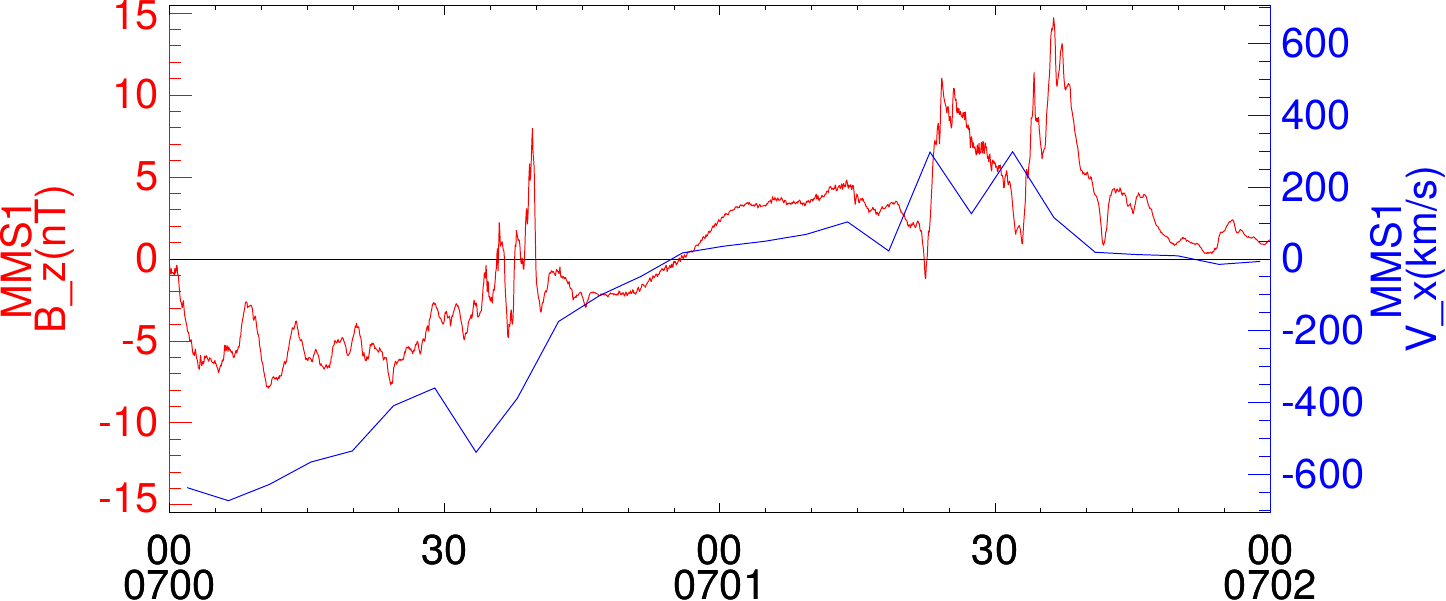}
	\caption{Components of the ion velocity moment and magnetic field measured by MMS1 on 26 July, 2017.  The correlated reversal of the ion velocity in the GSM$_{x}$ direction (blue) and the normal (GSM$_{z}$) component of the magnetic field (red) as observed by MMS1.  Given the small separation in time between the velocity and magnetic field reversals and the magnitudes of each both before and after the reversals, this example satisfies the algorithmic requirements for Stage 1.}
	\label{fig:S1-plot}
\end{figure}

Stage 1 consisted of searching for correlated reversals of the $\hat{x}$-component of the ion flow and the $\hat{z}$ component of the magnetic field within the segment.  Flow reversals are further required to be of at least $200 \frac{km}{s}$ in magnitude and B$_{z}$ reversals are required to be of at least $2nT$ in magnitude centered about zero.  Correlation of these reversals is determined by requiring that V$_{x}$ should turn from positive (negative) to negative (positive) within 90 seconds of B$_{z}$ turning from positive (negative) to negative (positive).  Segments which satisfy these criteria are then analyzed using the criteria for stage 2.  Figure \ref{fig:S1-plot} shows an example of a segment which passes the Stage 1 criteria where $B_{z}$ and $V_{x}$ are represented respectively by red and blue traces.  During these 2 minutes the flow changes from $V_{x} < -400\frac{km}{s}$ to $\sim 200 \frac{km}{s}$ (i.e. tailward to sunward) and, at practically the same time, $B_{z}$ goes from $-7 nT$ to $4 nT$ (i.e. southward to northward-pointing). This is consistent with a tailward motion of the X-line (Øieroset et al. 2001, Runov et al. 2003, Borg et al. 2005).

\begin{figure}[h!]
	\centering
	\includegraphics[width=\textwidth]{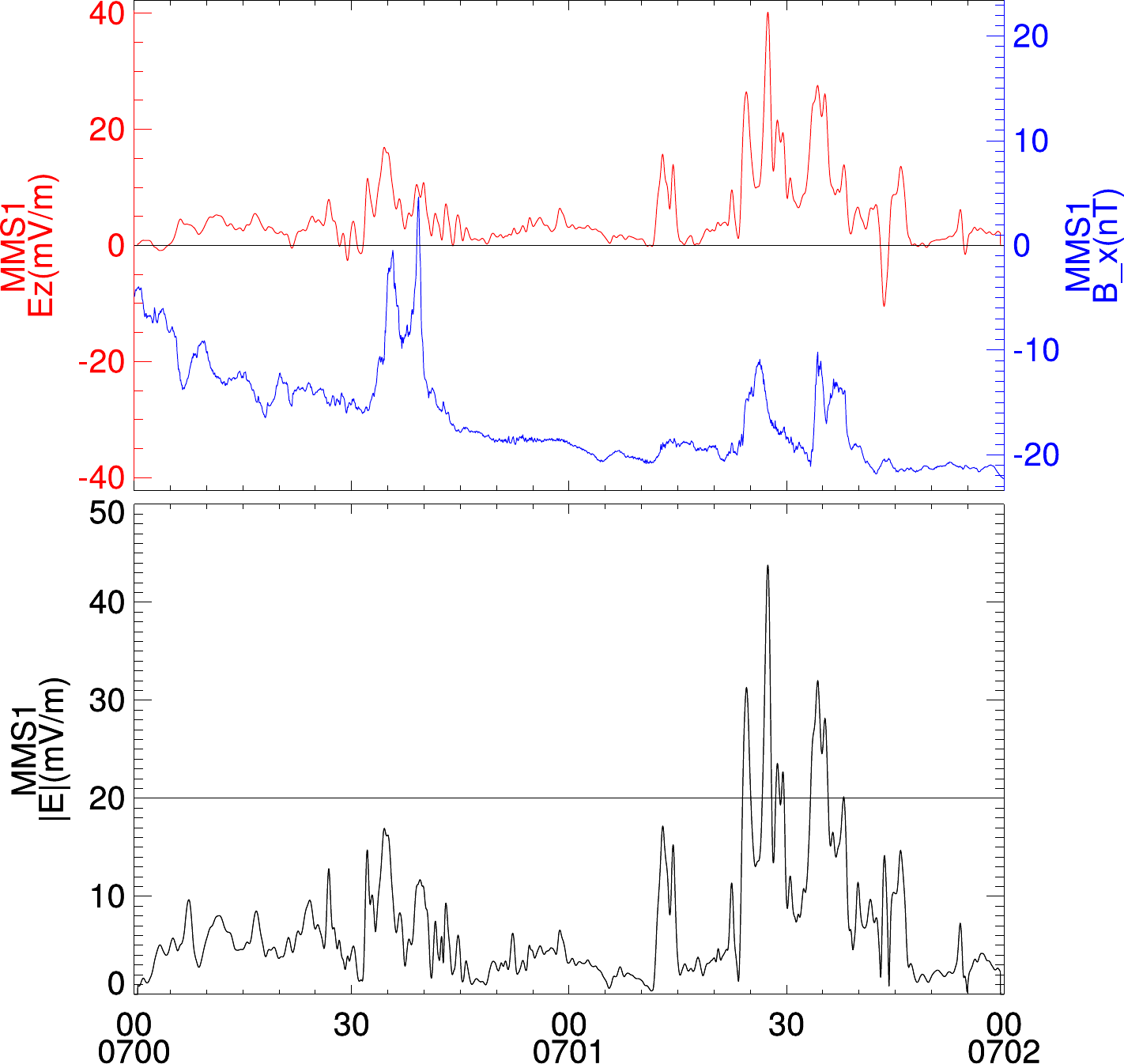}
	\caption{Components of the electric and magnetic fields measured by MMS1 on 26 July, 2017.  The magnitude of the electric field (top panel) is strong in the neighborhood (within 40s) of the correlated magnetic field and flow reversals as shown above.  We also note the strong GSM$_{z}$ electric field (in red, bottom panel) which is sign anti-correlated with the GSM$_{x}$ magnetic field component (in blue, bottom panel) which suggests the presence of the Hall electric field.}
	\label{fig:S2-plot}
\end{figure}

\begin{figure}[h!]
	\centering
	\includegraphics[width=\textwidth]{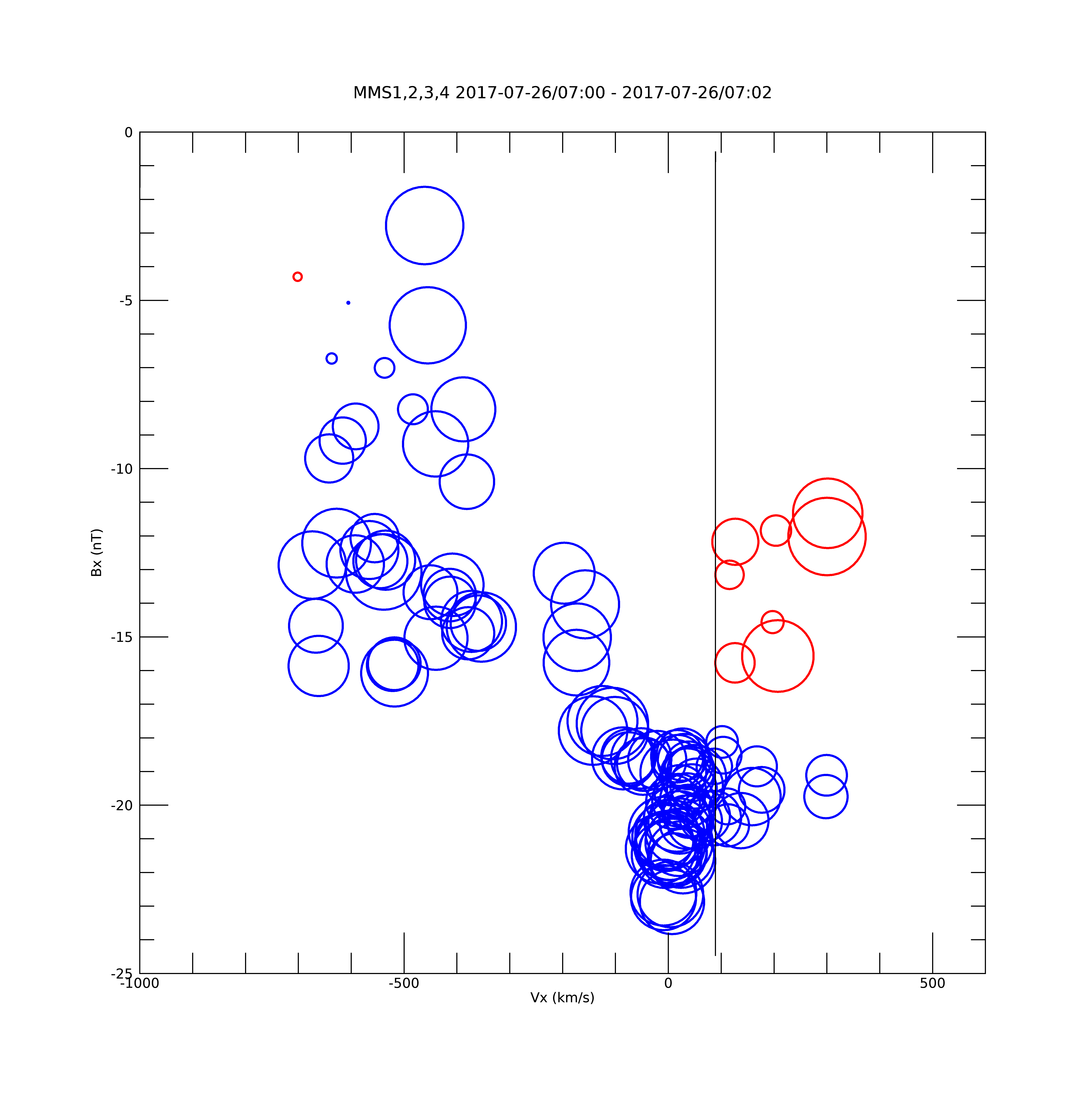}
	\caption{Bubbles represent the out-of-plane component (y) of the magnetic field, where blue circles are positive values and red are negative with area proportional to the square root of the magnitude of B$_{y}$ $\in$ $[0.1, 5.6] nT$, after Borg et al. (2005).  The $\hat{x}$ component of the measured ion velocity is the abscissa of the plot, while the ordinate axis is the $\hat{x}$ component of the magnetic field, providing a proxy coordinate system centered on the X-line.  Two quadrants of the quadrupolar Hall magnetic field are visible here, with moderate cross-contamination near the $V_{i}=0$ point.}
	\label{fig:S3-plot}
\end{figure}

Stage 2 consists of a search for Hall electric and magnetic fields within the segment.  Hall electric fields are determined by measuring the $\hat{z}$-component of the electric field with a magnitude of $\geq 3\frac{mV}{m}$ sign anti-correlated with the $\hat{x}$-component of the magnetic field, both in GSM coordinates.  The top panel of Figure \ref{fig:S2-plot} shows $B_{x}$ (blue trace) and $E_{z}$ (red trace) for the same time period as in Figure \ref{S1-plot}.  $B_{x}$ is negative throughout, indicating the spacecraft is south of the center of the current sheet. Except for brief excursions, $E_{z}$, which is normal to the mid-tail current sheet, is positive.  Thus it is directed towards the current sheet, as expected from the Hall Electric field.  All Hall magnetic fields are determined by measuring the GSM$_{y}$ component of the magnetic field $\geq 1nT$ with sign equal to $\operatorname{Sign}(B_{x})\times\operatorname{Sign}(V_{x})$ as shown in Figure \ref{fig:xline}. At least two of the four quadrants implied by this coordinate system, marked with Roman numerals in Figure \ref{fig:xline}, must be populated by B$_{y}$ measurements of the correct sign and magnitude for them to be accepted as evidence of the presence of Hall $\vec{B}$-fields.  Figure \ref{fig:S3-plot} illustrates this.  The spacecraft is sampling below the neutral sheet (negative $B_{x}$).  Tailward of the X-line ($V_{x}$ negative) the $B_{y}$ component is positive (blue bubbles) and earthward of the X-line it is negative (red bubbles). There is a tendency for the magnitude of the Hall $\vec{B}$ field to be greater close to the neutral sheet and the center of the reconnection region. 

Stage 3 consists of detection of a strong DC electric field of magnitude $\geq 10 \frac{mV}{m}$ in the neighborhood of the correlated B$_{z}$ and V$_{x}$ reversal and after having shown good evidence of Hall electric and magnetic fields.  To limit the influence of strong wave activity in this stage, the electric field data is low-pass filtered with a critical frequency of $1Hz$ prior to analysis.  Figure \ref{fig:S2-plot} shows an example of this on the same event as presented in Figure \ref{fig:S1-plot} where |$\vec{E}$| (lower graph) reached values greater than 40 $\frac{mV}{m}$ shortly after the correlated reversals at 0701UT.

If the segment satisfies these criteria, having already satisfied all previous criteria, then it is considered to contain a candidate IDR.

\subsection{Algorithm Implementation}

The algorithm was implemented through an IDL procedure, attached in the supplementary material for this article and available from a github repository (https://github.com/unh-mms-rogers/IDR\_tail\_search), leveraging the SPEDAS library for loading and version control of all MMS data files.  The procedure checks for each of the stages described above by way of conditional Boolean statements applied to each three-minute time series data segment.  Results are returned in the form of a text file which lists time segments which pass stage 1 criteria as well as markers indicating if the segment has also passed stage 2, or both stages 2 and 3.  Segments which have passed all three automated stages are then reviewed by a human researcher.  IDRs are only reported and confirmed after human review.   

All electric field data was processed using a low-pass digital filter with a critical frequency of 1 Hz prior to application of the algorithm in order to reduce false positives due to wave activity during the stage 3 analysis.  Fast survey data was used for this analysis; further investigation can and should be done using burst-mode data where available.  FPI measurements of particle moments were used for all reported IDRs and are the default source for particle moments.  The implementation of the above algorithm automatically falls back onto using HPCA proton moments when FPI moments are not available, although that was not necessary for any of the selected events.

Hall magnetic fields are tested by quadrant (see Figure \ref{fig:xline} for quadrant identification).  Each B$_{y}$ data point is checked for the minimum magnitude and correct polarity required to satisfy expected Hall magnetic field parameters in that quadrant.  After checking all B$_{y}$ data in the 3-minute time segment under analysis, a ratio of data points which satisfy expected Hall parameters to the total data points in the segment is calculated.  This is compared to a minimum threshold ratio entered as a user-defined parameter at runtime.  If the 'good' B$_{y}$ data points exceed the threshold ratio then the time segment is marked as having signatures of a likely Hall magnetic field.  The ratio used for this study was 0.10.

Final examination of code-identified IDR candidates is performed by eye using survey plots similar to those shown in Figure \ref{fig:s3a-example} and Figure \ref{fig:s3b-example}.  Human review is focused on checking for clear flow and magnetic field reversals with a minimum of erratic or noisy behavior which may call into question the timing or certainty of the reversals.  The electric field in the neighborhood of the reversal is also reviewed to ensure that strong wave activity does not mimic a DC field of sufficient size to pass Stage 3 even after the low-pass filter has been applied.  Hall electric and magnetic fields are reviewed for non-Hall behavior near the magnetic field and flow reversal.  Hall magnetic fields are checked using bubble plots similar to Figure \ref{fig:S3-plot}.

Non-ideal behavior in any of the three stages can still pass checks in the algorithm as currently implemented and may represent other geospatial events which are not ion diffusion regions (see Discussion).  These events are currently removed from consideration after review of the candidates by a human reviewer.  Modifications to the algorithm and the code which applies it are currently under development to further reduce the need and extent of human review of IDR candidates, as well as to possibly automate confident identification of dipolarization fronts, bursty bulk flows, non-active flow reversals, etc. for later study.

\section{Examples}

We now offer two case studies to illustrate the working of the algorithm.

\subsection{Case study 1: 0729UT July 26, 2017}
\begin{figure}[h!]
	\centering
	\includegraphics[width=35pc]{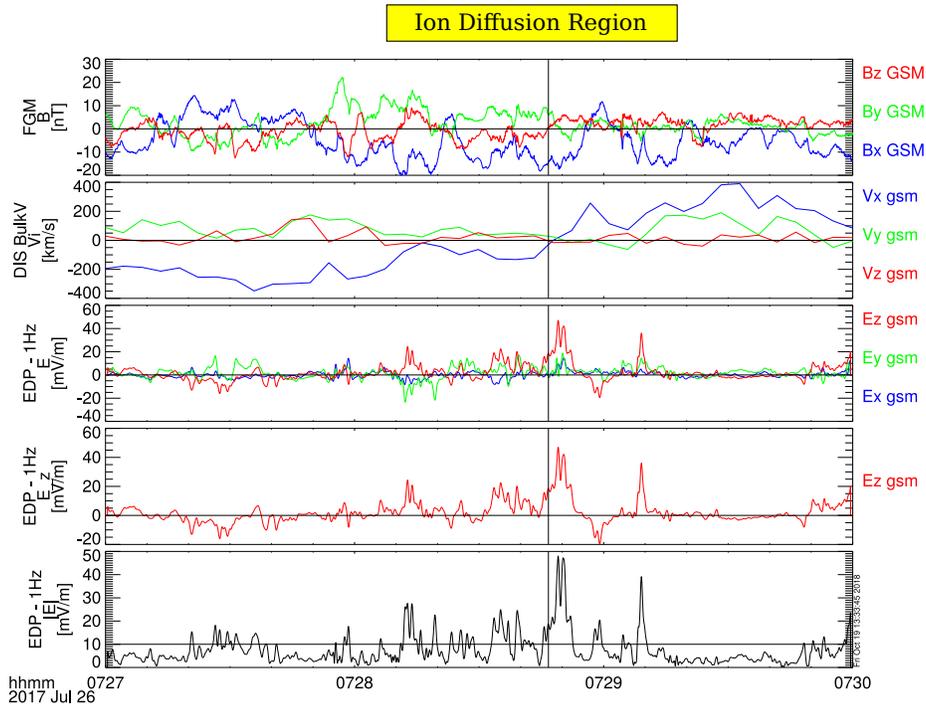}
	\caption{A strong reversal in B$_{z}$ (vertical line) is seen $\sim$3s before a correlated reversal in V$_{x}$.  The 30s surrounding these reversals also have strong electric fields, frequently much greater than $20\frac{mV}{m}$.  Other regions with strong electric field indicators of reconnection occur both before and after the correlated reversals (07:28:12, 07:30:00).  Hall electric fields were also measured in each of these regions and Hall magnetic fields are seen throughout the neighborhood surrounding the reversals.  Based on these indicators the time period where the observatory is within the diffusion region is marked by the colored label at the top of the Figure.}
	\label{fig:s3a-example}
\end{figure}

\begin{figure}[h!]
	\centering
	\includegraphics[width=\textwidth]{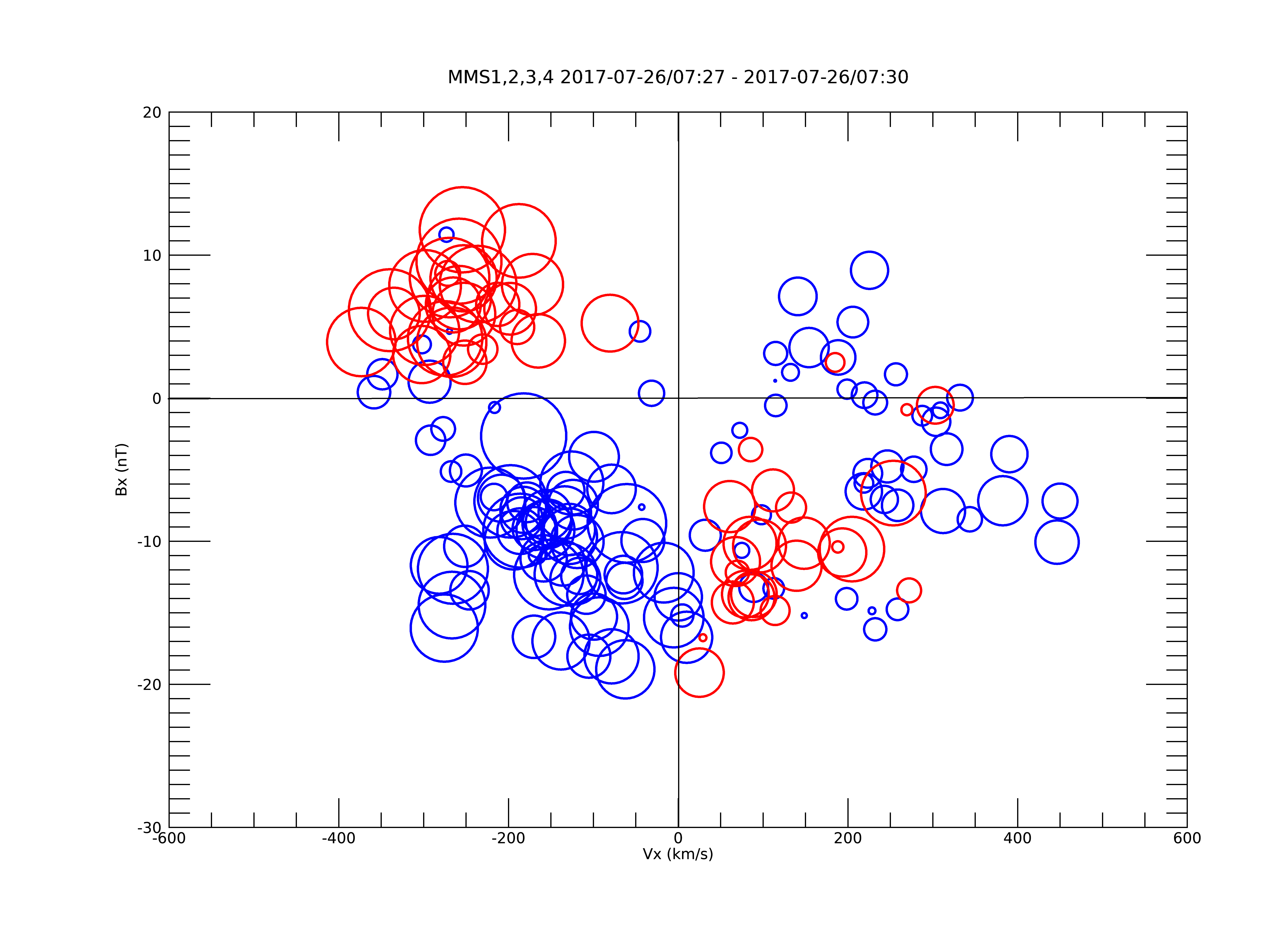}
	\caption{In this example all four quadrants of the Hall magnetic field (GSM B$_{y}$)are detected by the MMS observatories.  The time period of data points used in this plot is the same as for the time series data shown in Figure 5.  Colors are the same as in Figure 4, with the area of each circle again proportional to the square root of the magnitude of B$_{y}$.}
	\label{fig:s3a-bubble}
\end{figure}

Figure \ref{fig:s3a-example} shows three minutes surrounding an ion diffusion region observed by the MMS4 observatory as identified by the algorithm described in the previous section. This event was previously studied by Ergun et al. (2018) whose emphasis was on the turbulent energy transfer processes occurring in the period between 07:21:13UT and 07:38:42UT.  The location of MMS was (-23.0, 8.94, 2.20)$R_{E}$ in GSM coordinates.  The fluxgate magnetometer data in the top panel shows a rapidly changing  magnetic field with no fewer than 10 current sheet crossings (reversal in the polarity of B$_{x}$) in the period shown.  The z-component of the magnetic field also changes sign several times, with a particular reversal of interest from southward to northward at 07:28:46.7UT and marked with a vertical guideline.  In the second panel, ion speeds up to $400 \frac{km}{s}$ are recorded, with a transition from tailward to earthward flow direction at 07:28:48UT.   Filtered electric field data in GSM coordinates is shown, with additional panels dedicated to the z-component of $\vec{E}$ and to the magnitude of the electric field.  Strong DC electric fields ($\geq 10\frac{mV}{m}$) are measured frequently during the minute surrounding the correlated ion flow and B$_{z}$ reversal, reaching $\sim$40 mV/m during the reversal.  Comparison of E$_{z}$ and B$_{x}$ show the conditions expected for Hall electric fields in this region (marked in yellow). 

Figure \ref{fig:s3a-bubble} shows the y-component (out-of-plane) of the magnetic field ordered by the x-components of both the magnetic field and ion velocity.  Three quadrants of the Hall magnetic field are clearly observed by the combined measurements of all four MMS observatories.  All four quadrants are observed, although quadrant II, as described in Figure 1, has only sparse coverage.  The out-of-plane magnetic fields are stronger tailward of the X-line, both north and south of the neutral sheet; a trend typical of events in this study.

\subsection{Case study 2: 0749UT July 17, 2017}
\begin{figure}[h!]
	\centering
	\includegraphics[width=\textwidth]{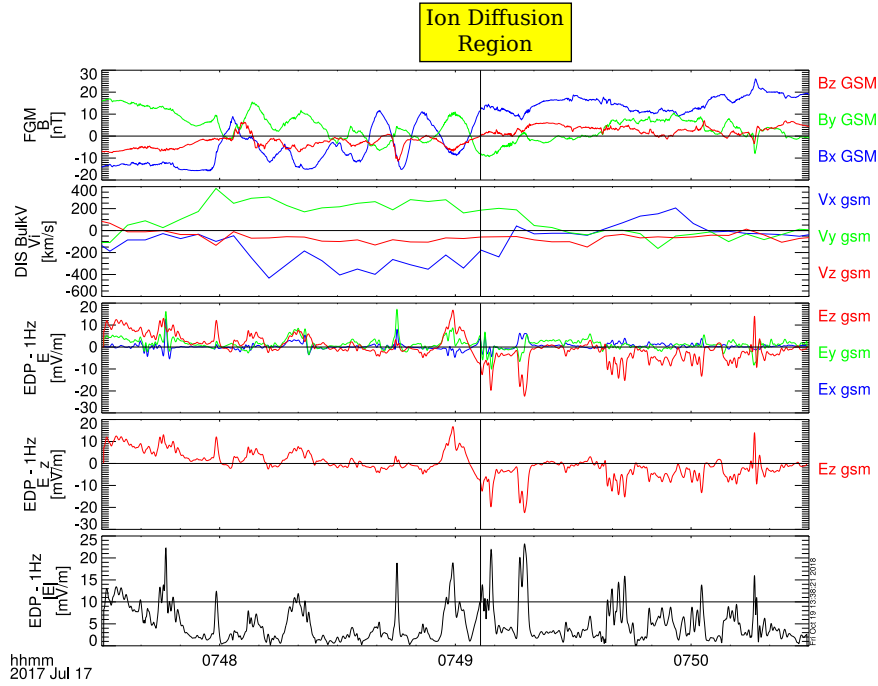}
	\caption{A moderate reversal in B$_{z}$ from $-11nT$ to $6nT$ is the last of a rapid series of neutral sheet crossings, with an associated ion flow reversal approximately 10s later.  Moderate to strong Hall electric fields are seen immediately surrounding the reversal.}
	\label{fig:s3b-example}
\end{figure}

\begin{figure}[h!]
	\centering
	\includegraphics[width=\textwidth]{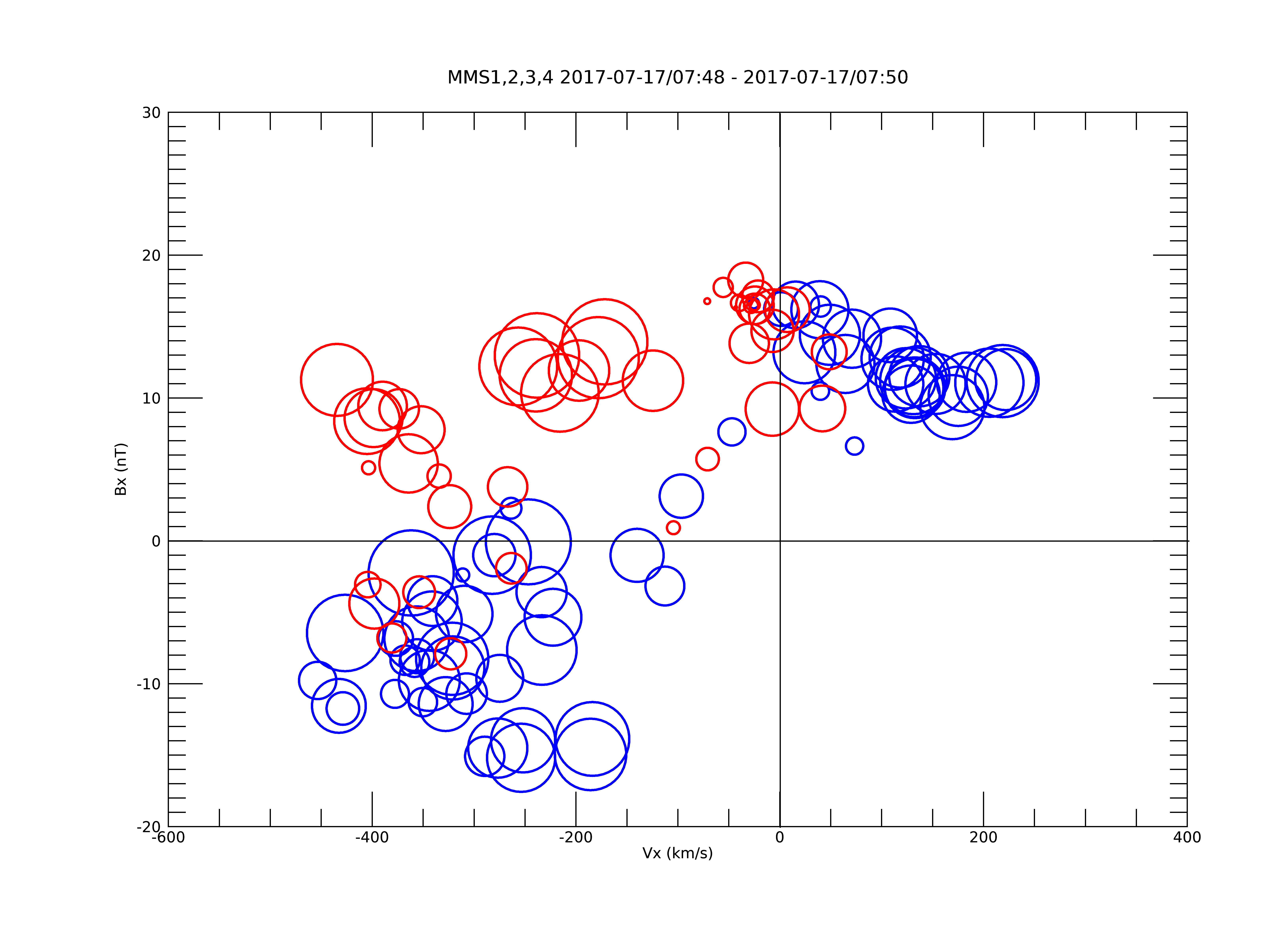}
	\caption{In this example, showing the same time period as in Figure \ref{fig:s3b-example}, three quadrants of the Hall magnetic field were observed by the MMS fleet in the minute surrounding the B$_{z}$ reversal.  The organisation of the measured Hall magnetic fields implies a small structure velocity for the X-line and surrounding diffusion region.}
	\label{fig:s3b-bubble}
\end{figure}

	Figure \ref{fig:s3b-example} shows time series data for three minutes of observations by the MMS3 observatory, roughly centered on an ion diffusion region as identified by our proposed algorithm.  The location of MMS was (-18.1, 7.30, 0.66)$R_{E}$ in GSM coordinates, 22.5 MLT at a distance of 19.5 $R_{E}$, therefore also on the dusk side.  The magnetic field data (top panel) shows seven suspected current sheet crossings during this period, the final crossing occurring shortly before the zero-crossing of interest in B$_{z}$ at 07:49:06.4UT.  Ion velocity moment data is given in the second panel and shows strong tailward flows of approximately $400\frac{km}{s}$ throughout the region of current sheet crossings before reversing to weaker earthward flows of about $200\frac{km}{s}$  approximately ten seconds after the correlated B$_{z}$ reversal.  Filtered electric field data is provided showing moderate-strength electric fields in the neighborhood of and in the minute following the correlated magnetic field and ion flow reversals.  Comparison of the z-component of the electric field with the x-component of the magnetic field indicates the presence of Hall electric fields in the seconds before and the minute following the correlated magnetic field and ion flow reversals.  Figure \ref{fig:s3b-bubble} shows the y-component (out-of-plane component) of the magnetic field in terms of B$_{x}$ and V$_{i,x}$ where three quadrants of the Hall magnetic field are clearly evident.

\section{Output of the 3-step selection scheme}
\begin{table}[h!]
	\normalsize
	\centering
	\begin{tabular}{|c|c|} \hline
		Stage Passed & \# of Events \\ \hline \hline
		Stage 1 & 148 \\ \hline
		Stage 2 & 37 \\ \hline
		Stage 3 & 17 \\ \hline
	\end{tabular}
	\caption{A table listing the stage of analysis and the number of events in the range from May 01, 2017 to September 30, 2017 which passed each stage.}
	\label{tab:selection-stats}
\end{table}

\begin{table}[h!]
	\normalsize
	\centering
	\begin{tabular}{|c|c|c|c|c|c|}  \hline
		Event Label & yyyy-mm-dd/tttt UT  & X-Line Motion & GSM$_{X}$(R$_{E}$) & GSM$_{Y}$(R$_{E}$) & GSM$_{Z}$(R$_{E}$) \\ \hline
		A & 2017-05-28/0358  & Tailward  & -19.3 & -11.8 & 0.78\\ \hline
		B & 2017-07-03/0527  & Tailward  & -17.6 &  3.32 & 1.75\\ \hline
		C & 2017-07-06/1535  & Tailward  & -24.1 &  1.41 & 4.44 \\ \hline
		D & 2017-07-06/1546  & Tailward  & -24.2 &  1.33 & 4.48 \\ \hline
		E$^{a}$ & 2017-07-11/2234  & Tailward  & -21.5 & 4.12 & 3.78 \\ \hline
		F & 2017-07-17/0749  & Tailward  & -18.1 & 7.30 & 0.66 \\ \hline
		G & 2017-07-26/0003  & Tailward  & -20.7 & 9.05 & 3.46 \\ \hline
		H & 2017-07-26/0701  & Tailward  & -22.9 & 8.97 & 2.27 \\ \hline
		I$^{b}$ & 2017-07-26/0728  & Tailward  & -23.0 & 8.94 & 2.20 \\ \hline
		J & 2017-08-06/0514  & Tailward  & -18.9 & 13.0 & 0.37 \\ \hline
		K & 2017-08-07/1538  & Tailward  & -16.4 & 4.38 & 3.77 \\ \hline
		L & 2017-08-23/1753  & Earthward & -18.8 & 16.1 & 1.11 \\ \hline
	\end{tabular}
	\caption{The 12 IDRs identified by the algorithm and confirmed by human review. Event E (indicated by an 'a' superscript) has been previously reported by Torbert et al.(2018). Event I (indicated by a 'b' super script) has been previously reported by Ergun et al. (2018)}  
	\label{tab:event-list}
\end{table}

Table \ref{tab:selection-stats} shows the total number of events which pass each stage.  Events which pass stage one but not subsequent stages include examples of non-active flow reversals and other phenomena.  Events which also passed stage 2 exhibit both clear correlated B$_{z}$ and V$_{x}$ reversals as well as good evidence of Hall electric and magnetic fields, but weak electric field magnitude overall. Examples of these can be found in the discussion.  A table of all time segments in our study which passed stage 1 with markers for if the segment passed stage 2  or stage 3 is given in the supplementary materials.  

The twelve IDRs identified by the algorithm described in the previous section and confirmed on review are listed in table \ref{tab:event-list}.  Epochs given in column 2 refer to the approximate center of the identified IDR.  The direction of X-line motion is determined by the observed direction of ion flow reversal with earthward motion indicated by an initial earthward ion flow converting to a tailward flow.   The order of encountered ion flows is inverted for tailward motion of the X-line (Eastwood et al., 2010 and references therein). 

\begin{table}[h!]
	\centering
	\begin{tabular}{|c|c|c|c|c|}  \hline
		Event Label & yyyy-mm-dd/tttt UT  & GSM$_{X}$(R$_{E}$) & GSM$_{Y}$(R$_{E}$) & GSM$_{Z}$(R$_{E}$) \\ \hline
		N1 & 2017-07-03/0546 & -17.8 & 3.28 & 1.78\\ \hline
		N2 & 2017-07-16/0638 & -15.6 & -2.04 & 5.51\\ \hline
		N3 & 2017-07-23/0753 & -22.2 & 8.50 & 1.77\\ \hline
		N4 & 2017-08-04/0923 & -21.6 & 8.44 & 2.59\\ \hline
		N5 & 2017-08-31/1153 & -13.0 & 18.7 & -5.77\\ \hline
	\end{tabular}
	\caption{Five events which satisfied all three stages of the algorithm described in this paper but could not be confirmed to be Ion Diffusion Regions upon human review.}
	\label{tab:weak-list}
\end{table}
	
Five other events which passed all three stages of the automated analysis but which are not reported here as IDRs display periods within a 3-minute selection segment which might indicate an IDR, but which lack consistency.  These events are tabulated in table \ref{tab:weak-list} and discussed in a following section. 

\section{Statistical Results}

\begin{figure}[h!]
	\centering
        \includegraphics[width=0.8\linewidth]{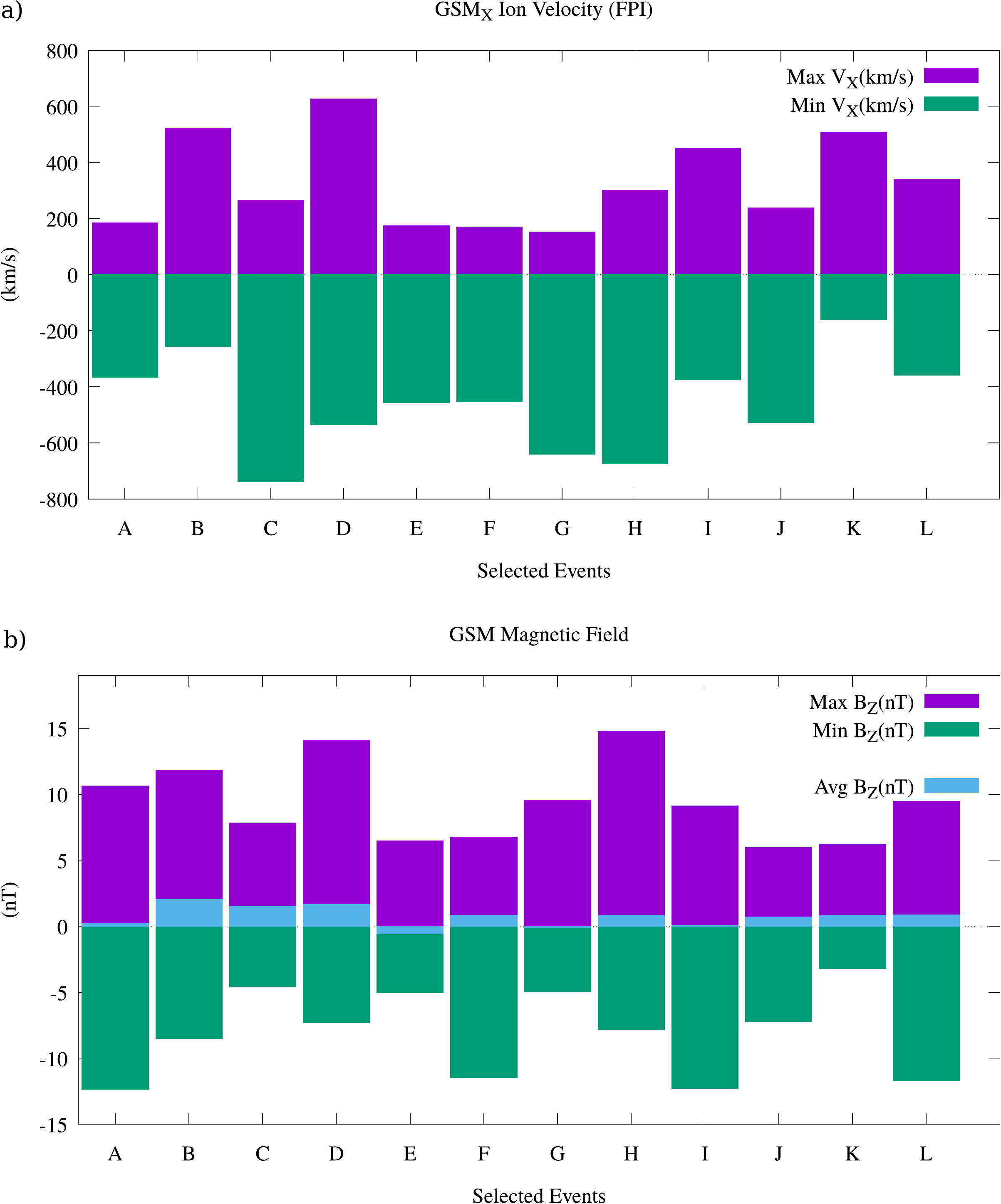}
	\caption{\underline{Stage 1 Criteria:}The above plots show the statistical properties of $V_{i,X}$ and $B_{Z}$ components in the neighborhood around the reconnection site.  Asymmetries in the flow offsets may indicate significant structure velocity.  Similar asymmetries or a non-zero average value of $B_{Z}$ are likely artefacts of the coordinates system not being boundary normal to the structure.  Comparison data from the Eastwood et al.(2010) study were not available.}
	\label{fig:S1-stats}
\end{figure}

Maximum positive and maximum negative values of the X-component of the ion bulk velocities for each of the nine reported events are shown in Figure \ref{fig:S1-stats} (top panel).  These values represent the extrema measured across all four spacecraft of the MMS fleet during the 3-minute time step approximately centered on each event.  No attempt has been made in these measurements to correct for structure motion relative to the spacecraft, thus asymmetries in the minimum and maximum may indicate tailward or earthward motion of the X-line relative to the spacecraft.  Any attempt to infer the magnitude of the X-line velocity using these asymmetries is hampered by the lack of a boundary-normal coordinate system.  Maximum negative (maximum pointing in a southward direction), maximum positive (maximum northward), and average values for the normal magnetic field are given in Figure \ref{fig:S1-stats}b (bottom panel).

\begin{figure}[h!]
	\centering
	\includegraphics[width=0.8\linewidth]{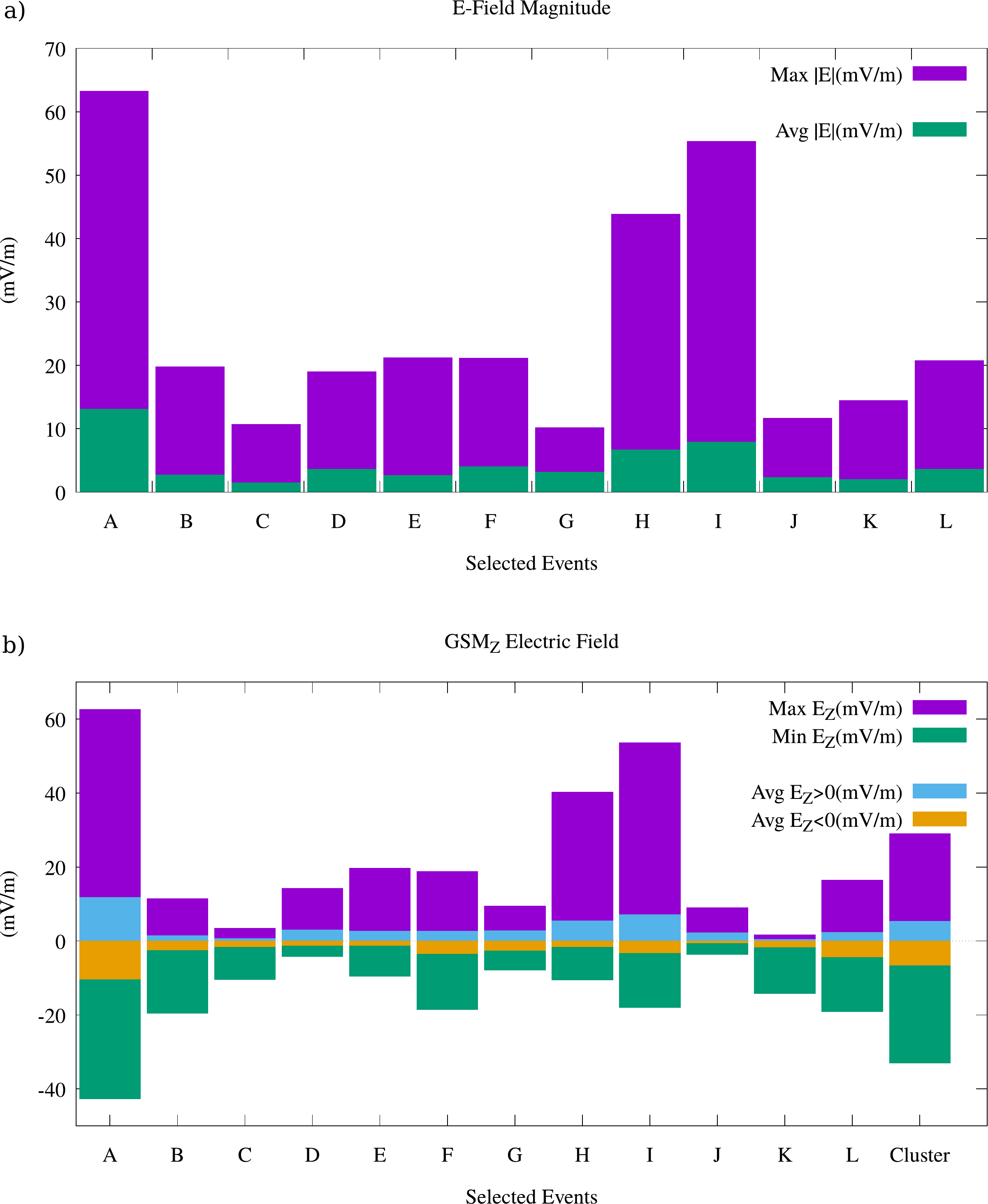}
	\caption{\underline{Stage 2 and Electric Field Criteria:} The upper plot, a),  shows statistical properties of the electric field magnitude identified IDRs in this study.  Data were not available from the Eastwood, et al.(2010) study to compare |$\vec{E}$|.  |E| $\geq$ 10$\frac{mV}{m}$ in the neighborhood surrounding the magnetic field and ion flow reversal is the requirement for passing Stage 3.  Below, in b),  are statistical properties of the Hall electric field (E$_{z}$) with comparison data from Eastwood, et al.(2010).  Detection of Hall electric fields (E$_{z}$ sign anti-correlated with B$_{x}$) is one half of the Stage 2 criteria for IDR candidate identification.}
	\label{fig:S2-stats}
\end{figure}

The upper panel in Figure \ref{fig:S2-stats} shows the maximum and average values of the electric field magnitude in the neighborhood including the diffusion region for each IDR.  As seen in Figures \ref{fig:s3a-example} and \ref{fig:s3b-example} the electric field strength increases significantly when approaching the inner diffusion region and drops off rapidly further into the outer diffusion region.  The broad difference between maximum and average values of $|E|$ suggest the existence of strong electric fields only very near the thin current sheet, with more moderate values elsewhere.  The lower panel in Figure \ref{fig:S2-stats} shows the largest and average values of the GSM$_{z}$ electric field (E$_{Z}$) for both positive and negative values.  We use the GSM$_{z}$ electric field as an approximation of the Hall electric field within the diffusion region, i.e. normal to the current sheet.  The dominating source of asymmetries is likely the path of the observatory through the diffusion region.  For the majority of the events reported, the distance to the current sheet, as approximated by the magnitude of B$_{X}$, was not uniform for observations on either side during a given event.  Extrema and average values for all events reported by Eastwood, et al. (2010) are also included for comparison and are discussed more later.

\begin{figure}[h!]
	\centering
	\includegraphics[width=0.8\linewidth]{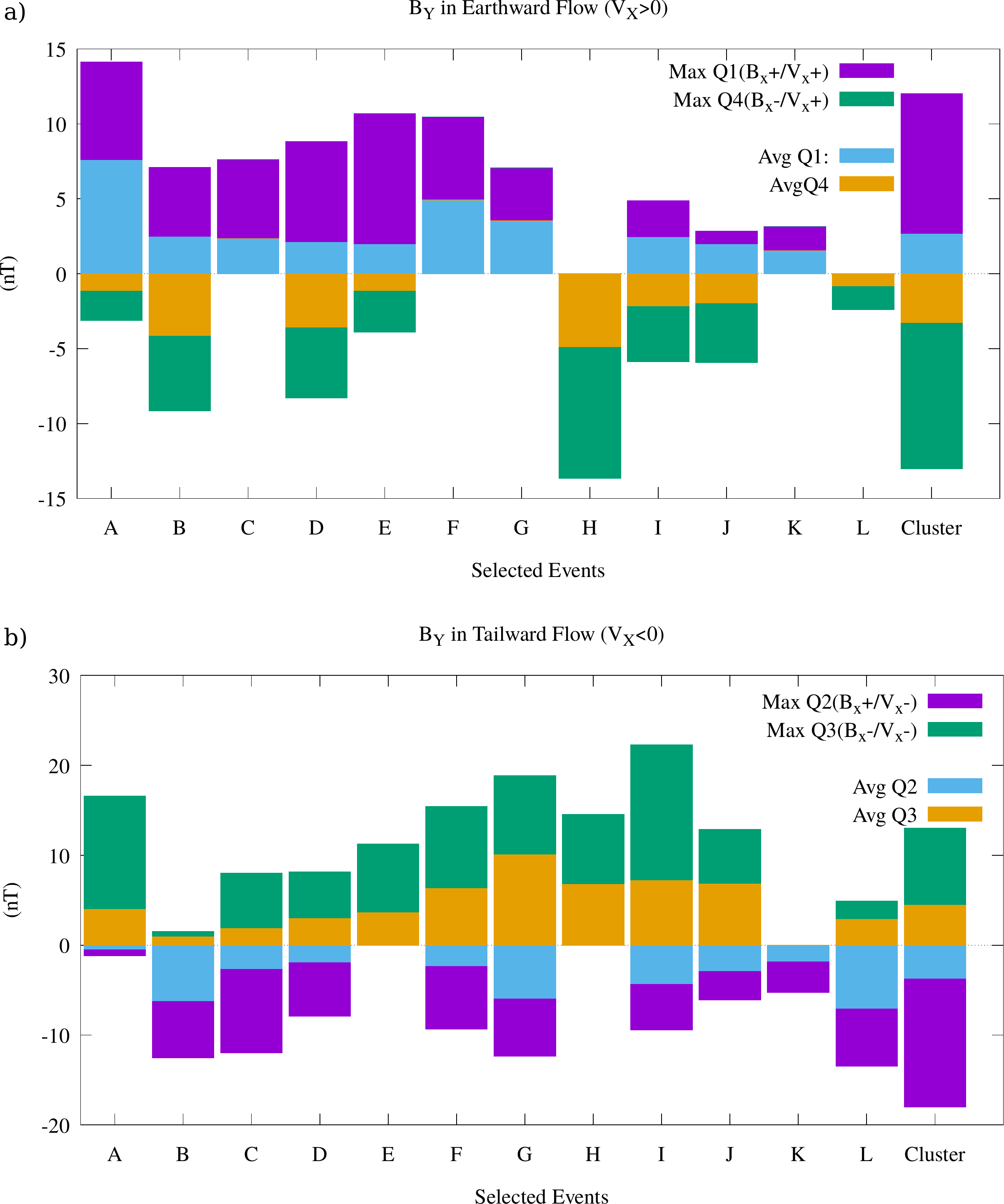}
	\caption{\underline{Out-Of-Plane (Hall) Magnetic Fields:} Detection of the quadrupolar Hall magnetic field is one half of the third stage in our IDR detection algorithm.  Properties of the out-of-plane magnetic field are divided into the four regions of the Hall model.  Not all quadrants were observed by MMS for some events.  Hall magnetic field data from the Eastwood et al.(2010) study are included for comparison.}
	\label{fig:S3-stats}
\end{figure}

Extreme and average values for all four quadrants of the Hall magnetic field, approximated by the out-of-plane magnetic field (B$_{Y}$), are given in Figure \ref{fig:S3-stats}.  These measurements are taken from across the entire selection window of 3 minutes. As with the Hall electric fields, the relative magnitude of Hall magnetic field seen in any given quadrant is dependent to a significant degree on the path of the observatory through the diffusion region.  Despite the small average separation between spacecraft in the fleet during this mission phase, we saw three and sometimes all four of the Hall magnetic field quadrants. Extrema and average values for all events reported by Eastwood, et al. (2010) are also included for comparison.  

The comparison data from Eastwood, et al.(2010) for E$_{z}$ and out-of-plane magnetic fields shows the maximum positive and maximum negative values across all 18 IDRs which are reported there, while the average positive and negative values are the averages of average values reported for each event.  Average values do not include events where the relevant field was not detected.

\begin{figure}[ht!]
	\centering
	\includegraphics[width=0.8\linewidth]{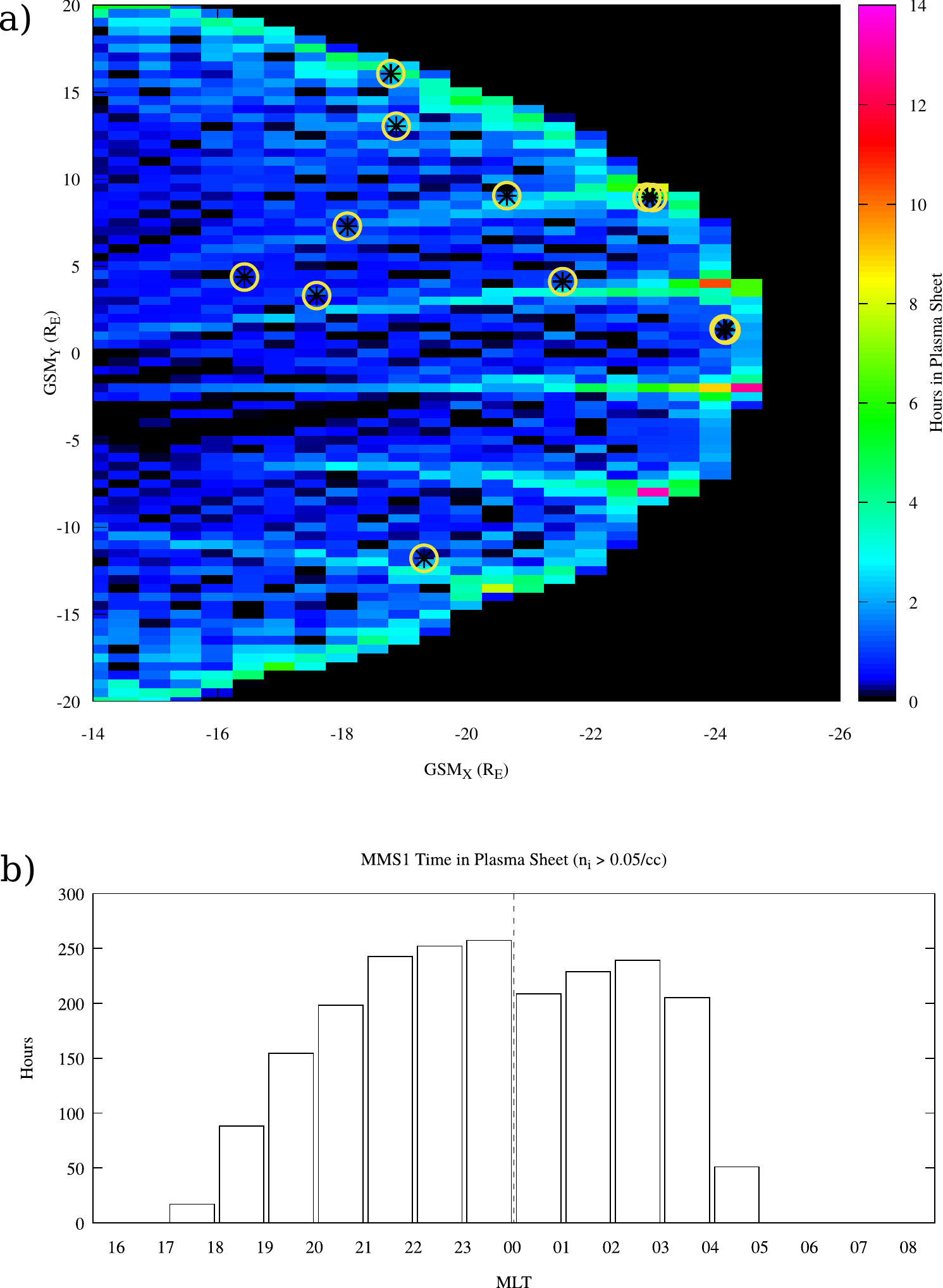}
	\caption{Locations of identified and confirmed IDRs in circled stars are plotted over the number of hours MMS spacecraft spent in the magnetotail.  The strong dusk-side prevalence of events is consistent with previous observations and predictions.}
	\label{fig:dwell-xy}
\end{figure}

Figure \ref{fig:dwell-xy}a) shows the distribution of IDRs as listed in table \ref{tab:event-list} over an underlay of the dwell time of the MMS fleet in the near-tail plasma sheet as determined by reported flight ephemeris and FPI ion density.  Figure \ref{fig:dwell-xy}b) shows the total hours spent in the plasma sheet by MMS as a function of position in MLT.  The event distribution shows a clear preference for the dusk region of the magnetotail (56.5\% dusk vs 43.5\% dawn).  The lone exception to this trend in our observations occurs on May 28, 2017 during a period when there was significant geomagnetic activity with an AE index in excess of 1000nT.

\section{Discussion}
\label{sec:disc}
\subsection{Mechanics and Limitations to the Algorithm}
The criteria employed in this algorithm are, in most respects, highly conservative when attempting to identify IDRs.  Our adherence to only considering those IDRs which are well aligned with the GSM coordinate system may eliminate many otherwise strong candidates.  Similarly, requiring the detection of an ion flow reversal is highly restrictive and eliminates candidates similar to the diffusion region encountered by Polar as described by Mozer, Bale, and Phan (2002) which would not have passed the first stage of analysis since the traversal was normal to the current sheet and on one side of the X-line. More qualitative criteria such as that of requiring smooth and rapid reversals of both ion flow and the normal component of the magnetic field also remove many events and conditions, such as a bifurcated flow reversal like that described by Runov et al.(2005), which might otherwise be argued to represent passage of MMS through the ion diffusion region. 

Our requirement of a strong $|\vec{E}| \ge 10\frac{mV}{m}$ also serves to limit detection of weaker or secondary reconnection diffusion regions such as those reported by Huang, et al. (2018) and Zhou et al. (2019).  These events can display weak but clear Hall electric and magnetic fields in the neighborhood surrounding a clear ion flow and normal magnetic field reversal, but are accompanied by only a small increase in the magnitude of the electric field (Zhou et al. 2019).  Such a weak electric field calls into question the existence of a thin current sheet in the region, a common feature of the canonical parallel reconnection model, but may still provide important or at least interesting examples.  Our criteria, both in terms of the electric field magnitude and the quality of the correlated ion flow and normal magnetic field reversal were made intentionally restrictive so as to, hopefully, provide a collection of examples of IDRs which can be considered such beyond a reasonable doubt.

\begin{figure}[ht!]
\centering
\includegraphics[width=\textwidth]{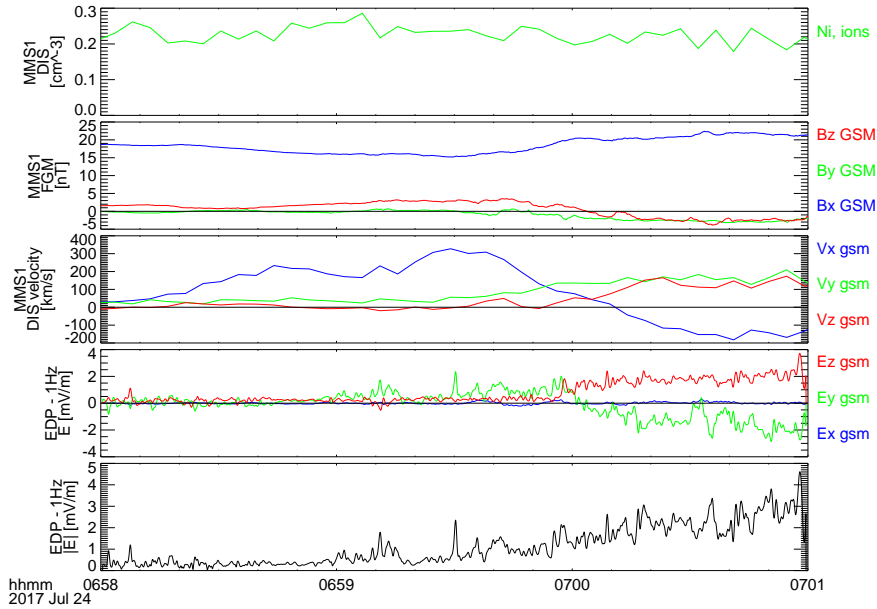}
	 \caption{This shows an ion flow reversal from earthward to tailward preceded by a correlated reversal in the GSM$_{z}$ component of the magnetic field $\approx$ 10s before, thus satisfying stage 1 criteria.  However other indicators of a possible IDR (Hall fields, strong |E|) are missing.  This is classified as a Non-Active Flow Reversal.}
 \label{fig:s1-example}
 \end{figure}

One noteworthy point regarding those flow reversals which do not display significant Hall fields is how common they are.  There is a factor of four drop in the number of events satisfying Stage 1 to those satisfying also Stage 2.  Almost half of the events which satisfied Stage 1 were identified as non-active flow reversals similar to those reported in Geotail observations by Nagai et al. (2013).  An example is given in Figure \ref{fig:s1-example}.  Here we see a clear correlated reversal from Earthward ion flow and north-ward magnetic field to a tailward flow and southward-pointing magnetic field.  However, strong electric fields are missing from the region, nor is significant evidence of Hall electric or magnetic fields found.  The missing elements combined with a steady, strong B$_{x}$, suggest passage of the observatory near to an X-line but at a distance from the current sheet such that no diffusion region was detected.

\begin{figure}[ht!]
	\centering
	\includegraphics[width=0.9\textwidth]{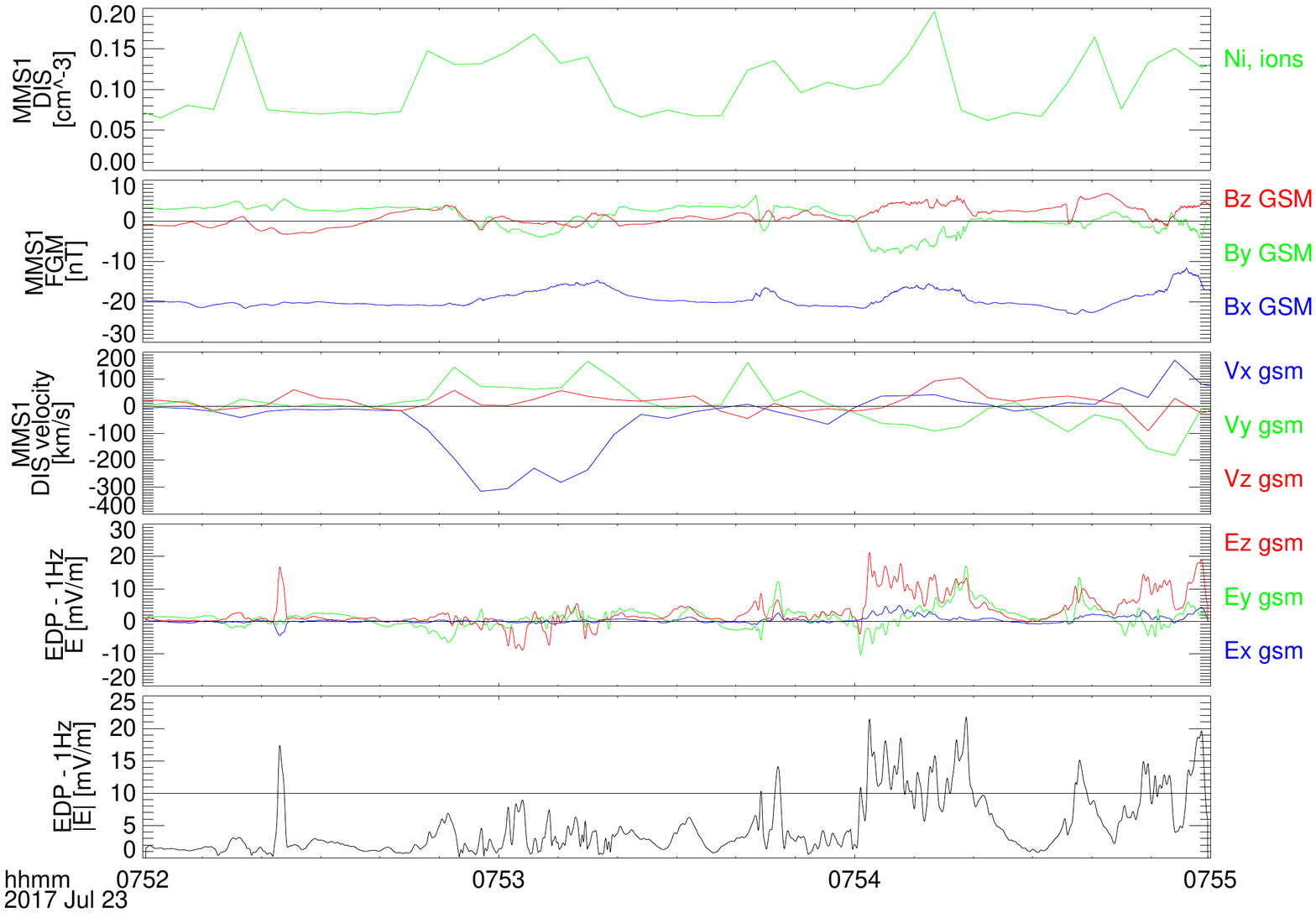}
	\caption{A survey plot of event N3.  This is an example of an interval which satisfied the criteria of all 3 stages, but which on visual inspection was not confirmed as an IDR.  Note the intermittent presence of Hall electric and magnetic fields and uncertain reversal in B$_{z}$.  Due to these conditions, and despite the strong, if bifurcated, ion flow reversal, the event was not determined to be an IDR}
	\label{fig:S3_noIDR}
\end{figure}

Some events which satisfy the criteria from all three stages are still questionable examples of IDRs.  An example of this is given in table \ref{tab:weak-list} as event N3 and shown in Figure \ref{fig:S3_noIDR}.  Here a strong ion flow reverses slowly from -315 to 170 $\frac{km}{s}$ in a moderately active magnetic field.  There are numerous instances during this reversal of B$_{z}$ crossing zero from negative to positive, but none appear to line up well with the flow reversal and both the closest in time and the most prominent B$_{Z}$ reversal occurs $\sim 50s$ before the flow reversal.  The Hall electric and magnetic fields are, likewise, indicated in places but are occasionally contradicted.  Despite strong electric fields consistent with a Hall electric field at the center of the bifurcated flow reversal, the electric fields elsewhere are less indicative of an IDR.  The flow reversal is strongly bifurcated with V$_{i} \sim 0$ for approximately one minute between distinct tailward and Earthward flows.  The strong, consistent magnitude of B$_{x}$ combined with the fairly weak plasma density raise the possibility that the observations were made in or near the plasma sheet boundary layer and away from the possible reconnection site. For these and similar reasons this event and three others which passed all three stages of the automated detection algorithm are not reported as IDRs.

Work is currently in progress to further refine the stages as they are described here and also to possibly include further requirements in an effort to ensure confidence in the identification made by the algorithm.  However, care must be taken to include as few assumptions as possible when considering refining criteria so as not to bias conclusions about the nature of the reconnection region based on the parameters used to find it.  

\subsection{Interpretation of Results}

The abundance of correlated magnetic field and ion flow reversals without significant electric fields or evidence of Hall fields calls into question a common assumption in the study of magnetic reconnection.  Non active flow reversals such as that shown in Figure \ref{fig:s1-example} represent a large portion of the total correlated field and flow reversals observed in this study.  This result goes against the received wisdom of correlated field and flow reversals being a sure indicator of a diffusion region, which has persisted for over a generation (Frank et al. 1976).  The inclusion of additional criteria, such as significant evidence of Hall electric and magnetic fields greatly improves the success of the algorithm, but still allows for some events which are not clearly diffusion regions and require further analysis before identification is certain.  

\subsection{Comparison with Cluster}

We now compare our observations with those made by Eastwood et al. (2010), wherein Cluster data from over a much longer period of time was analyzed.  The values for the proxy normal electric field fall within the range reported by Eastwood, et al.(2010) (their Figure 5) as seen in Figure \ref{fig:S2-stats}. Again, extreme and average out-of-plane magnetic field values for the events reported by Eastwood, et al. (2010) are also shown for comparison.  All four quadrants of the Hall magnetic field structure are observed clearly in $3/12$ events reported here with a typical spacecraft separation of $\sim 20km$.  The Cluster mission with spacecraft separation greater than $200km$ and often exceeding $1000km$ for all orbits reported by Eastwood, et al (2010) observed all four quadrants slightly more often, ($6/18$) diffusion regions, as expected from a larger spacecraft separation and the expected extent of Hall magnetic field signatures away from the diffusion region.  In both studies at least two and often three quadrants of the Hall magnetic field were observed for each event where the full quadrupolar was not. These factors along with the predominance of tailward over earthward motion of the X-line and the clear preference for dusk-side location suggest that the properties of the ion diffusion regions observed in this study are not fundamentally different than those observed during Cluster.

Several instances of normal electric field (E$_{z}$) in this study exceeded the maximum value reported for events observed by Cluster as reported by Eastwood, et al.  The reasons for this are unclear but may indicate a closer approach to the inner diffusion region than was done by the Cluster spacecraft during their encounters.  Certainly in the case of event I of this study the strong normal electric field coincided with an encounter with the inner diffusion region (Ergun, et al. 2018).  It may be that Cluster was not so fortunate during the events studied by Eastwood, et al.  The average E$_{z}$ across all twelve events reported here was $\sim 3.52\frac{mV}{m}$ for E$_{z}^{+}$ and $\sim -2.81\frac{mV}{m}$ for E$_{z}^{-}$ which, while smaller, compare reasonably well with $\sim5.33\frac{mV}{m}$ and $\sim -6.47\frac{mV}{m}$ for the same averages from the Eastwood, et al. study.

The average positive out-of-plane magnetic field observed in the twelve events reported here exceeded the average value of events reported in Eastwood, et al. in quadrants 1 ($\sim 3.08nT$ vs $\sim 2.64nT$) and 3 ($\sim 4.85nT$ vs $\sim 4.44nT$).  The average negative out-of-plane magnetic field is somewhat smaller than that reported by Eastwood, et al. in quadrant 4 ($\sim-2.46nT$ vs $\sim-3.25nT$) and quadrant 2 ($\sim 3.52nT$ vs $\sim 3.66nT$).  A significant source of variation in the magnitudes of out-of-plane magnetic fields is likely due to different flight paths through the diffusion region, which can vary greatly from one example to the next. In all cases the averages in B$_{y}$ magnitudes from each study are sufficiently similar to be confident that behavior of the events in this study are fundamentally similar to those reported by Eastwood, et al.

A notable trend is the greater magnitude of out-of-plane magnetic fields in regions with tailward flow (quadrants 2 and 3) relative to quadrants with Earthward flow (quadrants 1 and 4).  This trend is evident in both the events reported here as well as those reported by Eastwood, et al.  The ratio of average |B$_{y}$| for V$_{x} < 0$ to average |B$_{y}$| for V$_{x} > 0$ is $1.38$ for the events reported by Eastwood, et al. and $1.31$ for those reported here.  These are both of similar magnitude to the ratio of average peak tailward flow to average peak Earthward flow in events A--L of $\sim1.41$.

The direction of X-line motion was predominantly tailward except for one event (event 'L' on Table \ref{tab:event-list)} moving Earthward as indicated by the order of the ion flow reversal.  Similarly, Earthward moving events were in the minority during the Eastwood, et al. study with only $3/18$ Earthward-moving X-lines (Table 2, Eastwood, et al.(2010)).  There is a suggestion that some of these tailward moving X-lines may be near-Earth neutral lines moving tailward during the recovery phase of substorms.  Similarly, the Earthward-moving X-lines may be related to the expansion phase of a related substorm.  Further investigations regarding this question are currently underway.

\subsection{Further Discussion}

An interesting result on the statistics is the following: the total number of IDRs reported also conforms with predictions made by Genestreti et al. (2014) pre-launch, wherein a statistical analysis of the distribution of IDRs observed by Cluster and Geotail indicated that MMS ought to observe $11\pm 4$ IDRs during the first tail phase (Genestreti et al. 2014), i.e. the one covered in this study.  The twelve events we report is comfortably within this prediction.  The spatial distribution of events observed here strongly favors the dusk-side (GSM$_{y} > 0$) as was also predicted by Genestreti et al., based on the locations of IDRs previously observed by Cluster and Geotail, although the dawn-dusk asymmetry observed by MMS was far greater than that of previous missions (Figure 6, Genestreti et al. 2014).


During the 2017 MMS tail season the fleet spent a total of 2143.02 hrs in the plasma sheet ($n_{i} > 0.05 cc^{-1}$ by FPI fast survey).  Plasma sheet dwell time on the dusk side was 1210.30 hrs and on the dawn side 932.72 hrs.  This corresponds to a Dusk/Dawn ratio of 56.5\%/43.5\%.  Meanwhile, we observed 11 confirmed IDR events on the dusk side and one on the dawn side (91.7\%/8.3\%).  If the distribution of IDRs were to be even across the tail, we would instead expect to have seen 7 (5) events on the dusk (dawn) side.  We can, therefore, confidently state that the dawn-dusk asymmetry of reconnection events is not due to any observational bias caused by orbital variations.  A previous study by Raj et al. (2002) of high-speed ion flows (often associated with exhaust from reconnecting X-lines) in the geomagnetic tail also showed an asymmetry of events favoring the dusk side over the dawn side well in excess of the asymmetry in total number of measurements made in either region (their Figure 18).  Lu et al. (2018) propose a mechanism which produces a dawn-dusk asymmetry in the thickness of the tail current sheet and support it with PIC simulation results which approximate Cluster measurements.  The distribution of IDRs observed by MMS, however, is much greater than what is suggested in these previous studies.

\section{Concluding Remarks}

We have presented a numerical algorithm to identify IDRs and applied it to MMS \textit{in situ} observations in the geomagnetic tail during a 5-month period.  The algorithm uses a stepwise scheme, testing topological parameters in the magnetic and electric field as well as the bulk ion flow to allow a high degree of confidence in the regions identified by code.  Using this method we have identified 12 IDRs from the 2017 Phase 2B magnetotail campaign of the MMS mission.  After performing statistical analysis of these events and comparing them with previous surveys of magnetotail diffusion regions we find that the algorithmically identified IDRs have average properties which are  similar to those identified in previous studies.  We also demonstrate categorically that MMS has only a slight orbital dawn-dusk bias in its coverage of the plasma sheet (43.5\% - 56.5\%).  However this is much smaller than the asymmetry in the IDR observations (8.3\% - 91.7\%).  From this we conclude that the effect we show is a real bias in the occurrence of diffusion regions.

While our algorithm has been developed for and initially applied to MMS data, the criteria used are meant to be general for IDRs regardless of where they occur and are applicable without alteration to any mission in the geomagnetic tail.  The code, as currently implemented in IDL, requires minimal modification to be adapted for use on data from any other mission with support in the SPEDAS library.  The essential algorithm in the code could, with some small additional effort, be ported to other languages or adapted for use in other regions besides the tail.  The source code is made available, both on github and as a supplementary material here, for this purpose.



The efficacy of the algorithm is readily apparent in the results generated by its use and presented here in sections 4 and 5.  Analysis of those results, as presented in sections 6 and 7, show that the algorithm produces worthwhile events which are of interest to the community with a minimum of human intervention.  This presents a valuable, time-saving aid to researchers in the field of magnetic reconnection.

%
%

%

%


%

\acknowledgments
We thank the entire MMS team for the effort invested in the preparation of the data.  The authors would like to thank Kevin Gennestretti and Terry Forbes for useful conversations.
All MMS data are publicly available at MMS Science Data Center (https://lasp.colorado.edu/mms/sdc/public/).  Source code for the algorithm described is part of the supplementary materials for this article and is available from a github repository (https://github.com/unh-mms-rogers/IDR\_tail\_search).  
This work was supported by NASA contracts 499878Q,499935Q,NNX10AQ29G, NNX16AO04G, NNX15AB87G, NSF grant AGS-1435785.

\end{document}